\documentclass[preprint, amsmath, amssymb,aps,pre, superscriptaddress]{revtex4-2}
\usepackage{graphicx}
\usepackage{epstopdf}
\usepackage{dcolumn}
\usepackage{bm}
\usepackage{xcolor}
\usepackage{natbib}

\begin{document}

\title{The bouncing 
dynamics of inertial self-propelled 
particles \\
reveals directional asymmetry}

\author{Denis Horvath}
\affiliation{Center for Interdisciplinary Biosciences, Technology and Innovation Park, P. J. \v{S}af\'{a}rik University, Jesenn\'{a} 5, 041 54 Ko\v{s}ice, Slovak Republic}
\author{Cyril Slab\'{y}}
\affiliation{Department of Biophysics, Faculty of Science, P. J. \v{S}af\'{a}rik University, Jesenn\'{a} 5, 041 54 Ko\v{s}ice, Slovak Republic}  
\author{Zolt\'{a}n Tomori}
\affiliation{
Department of Biophysics,
Institute of Experimental Physics SAS, 
Watsonova 47, 040 01 Ko\v{s}ice, Slovak Republic}
\author{Andrej~Hovan}
\affiliation{Department of Biophysics, Faculty of Science, P. J. \v{S}af\'{a}rik University, Jesenn\'{a} 5, 041 54 Ko\v{s}ice, Slovak Republic}  
\author{Pavol Miskovsky}
\affiliation{Center for Interdisciplinary Biosciences, Technology and Innovation Park, P. J. \v{S}af\'{a}rik University, Jesenn\'{a} 5, 041 54 Ko\v{s}ice, Slovak Republic}
\author{Gregor B\'{a}n\'{o}}
\email{gregor.bano@upjs.sk}
\affiliation{Department of Biophysics, Faculty of Science, P. J. \v{S}af\'{a}rik University, Jesenn\'{a} 5, 041 54 Ko\v{s}ice, Slovak Republic}

\date{\today}

\begin{abstract}

This study aims to examine experimental conditions in which active particles are forced by their surroundings to move forward and backward in a continuous oscillatory manner. The experimental design is based on using a vibrating self-propelled toy-robot called hexbug, which is placed inside a narrow channel closed on one end by a rigid moving wall. Using the end-wall velocity as a controlling factor, the main forward mode of the hexbug movement can be turned to mostly rearward mode. We investigate the bouncing hexbug motion on both experimental and theoretical grounds. The Brownian model of active particles with inertia is employed in the theoretical framework. The model itself uses a pulsed Langevin equation in order to simulate abrupt changes in velocity that mimic hexbug propulsion in the moments when its legs make contact with the base plate. Significant directional asymmetry is caused by the legs bending backward. We demonstrate that the simulation successfully reproduces the experimental characteristics of hexbug motion after regressing the spatial and temporal statistical characteristics, especially when directional asymmetry is under consideration.
\end{abstract}


\maketitle

\section{\label{intro}Introduction}

Various phenomena arise in active matter as a result of the movement and interactions of self-propelled particles. The size of the systems under consideration ranges from microscopic self-propelled colloids \cite{Howse2007,Palacci2010,Palacci2013}, bacterial colonies \cite{DilEonardo2010, Cates2012, Nishiguchi2017} and artificial micro-swimmers \cite{Kaynak2017, Aghakhani2020}, through millimeter-sized vibrated granular particles \cite{Weber2013,Walsh2017, Dauchot2019} up to large-scale animal flocks or swarms \cite{Nagy2010, Lukeman2010, Attanasi2017}, as well as human crowds \cite{Bain2019, Bottinelli2016, Kulkarni2019}. Active matter exhibits a distinct non-thermal behavior that results from the propulsion of its constituent particles. In addition, particle interactions with the medium, with barriers and constraints, and with other particles result in a variety of collective effects.

The dynamics and collective behaviour of macroscopic (centimeter-sized) self-propelled particles moving autonomously on a 2D surface were studied intensively \cite{Giomi2013,Scholz2018a}. In most cases, propulsion was induced by a set of bent particle legs excited through vibrations, agitating either the table or the particles themselves \cite{Cicconofri2015, Koumakis2016}. As an alternative, vibrated granular particles were used in experiments \cite{Walsh2017,Weber2013,Murali2022}. Interestingly, all of these macroscopic self-propelled systems exhibited stochastic behavior similar to that of their microscopic equivalents. In fact, various modifications of the Active Brownian Particle model were successfully used to explain the emerging phenomena \cite{Weber2013,Scholz2018a}. However, unlike their microscopic counterparts, the inertia of the macroscale active particles was typically not negligible. Massive self-propelled particles showed an inertial delay between their orientation and velocity \cite{Scholz2018a}. On a parabolic landscape (antenna), particle inertia controlled the transition from the "orbiting" state to the "climbing" state  \cite{Dauchot2019} and affected the particle dynamics when moving near strongly confining boundaries \cite{Leoni2020}. Besides that, non-negligible particle inertia was found to influence the collective motion of active granular particles \cite{Scholz2018b, Loewen2020} and led to a coexistence between gaslike particle behaviour and surface clusters in circular arenas \cite{Deblais2018}. Finally, the effect of particle inertia was also studied in the theoretical framework of Active Ornstein-Uhlenbeck Particles \cite{Caprini2021, Nguyen2022}.

In this work we study the dynamics of a single self-propelled inertial particle bouncing against a wall. An individual hexbug toy-robot \cite{Dauchot2019,Leoni2020,Tapia2021} is used in the experiments. The hexbug is confined in a straight, narrow channel (see Fig.\ref{fig:hb}), which prevents rotation and maintains the effective propulsion force acting in one direction. A hard wall closes the channel end. The hexbug bounces off after striking this end-wall. A continual bouncing motion results from the propulsion pushing the hexbug back against the end-wall. Only the forward-backward motion is analyzed. When the end-wall is moved along the channel in either direction at a constant speed, an interesting phenomenon is observed. Plotting the hexbug mean distance from the end-wall against the speed of the wall reveals a dependence with a clear minimum. Our objective in this work is to provide an explanation for the observed characteristics.

The Active Brownian Particle model is modified to account for the specifics of the experiment. We introduce a pulsed version of the Langevin equation where the particle velocity varies discretely in steps at a given frequency. Inertial, viscous damping, propulsive and stochastic forces are taken into account. Besides that, the particle velocity changes as a result of collision with the hard wall at the channel end. The particle moves in a ballistic regime in the intervals between velocity-changing events, which is consistent with the experimental observations. To differentiate between the parameters used for particles moving forward and backward, directional asymmetry has been added to the model. This feature is found to be essential for explaining the experimental findings. The general consequences of this asymmetry are also examined to provide guidance for other similar systems.

\section{\label{methods}Materials and Methods}

\subsection{\label{experiment}Experimental}

\subsubsection{The experimental setup and conditions}

\begin{figure}[t]
\includegraphics[width=8cm]{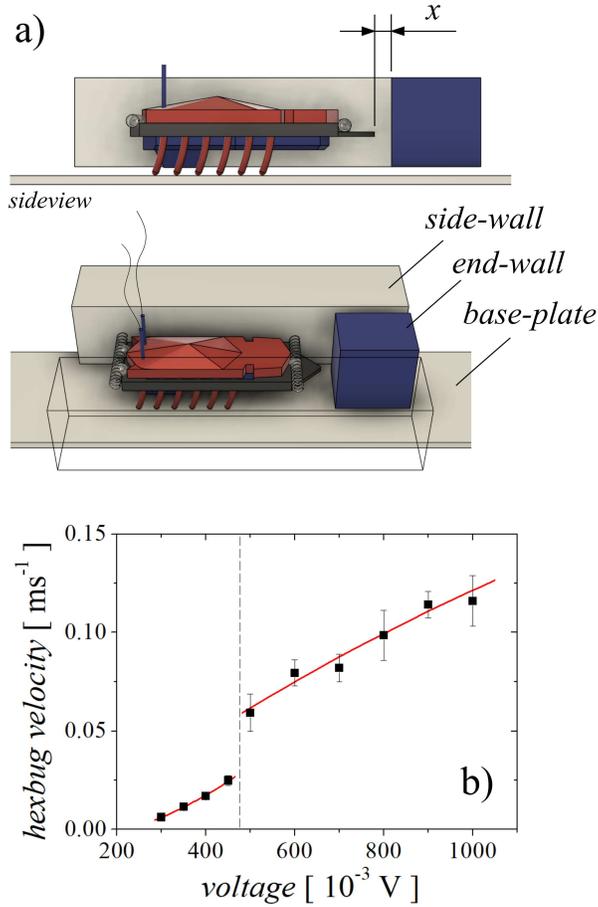}
\caption{\label{fig:hb} a) The modified hexbug inside the narrow channel closed by the end-wall. The 3D-printed plastic frame attached to the particle is indicated in black; $x$ defines the hexbug distance to the end-wall. b) The hexbug velocity in an open channel as measured for different voltages.}
\end{figure}

A modified hexbug toy-robot was used in the experiments (Fig.\ref{fig:hb}). The original twelve-leg robot, with dimensions: $length \times width \times height = 45{\rm mm}  \times 15{\rm mm} \times 15 {\rm mm}$ and mass of 7.5 g, is equipped with an internal vibrator. To avoid the battery getting low during the measurements, as observed e.g. in \cite{Dauchot2019}, the hexbug was powered through a pair of thin (100 micrometer diam.) copper wires connected to a dc power supply. The battery was removed from the hexbug body and was replaced by a weight of identical mass. The frequency of the vibrator was regulated by changing the power supply voltage.

The hexbug was also equipped with a 3D-printed outer plastic frame that was firmly glued to the hexbug body.  The frame was designed so as not to touch the hexbug legs. Our goal was to investigate only the forward-backward, quasi-one-dimensional hexbug motion in a narrow channel formed between two parallel side walls. A special effort was made to reduce any potential interference between the longitudinal and transversal particle dynamics. The pair of soft coil springs that were attached to the plastic frame and protruded on the sides of the hexbug gave the best results (see Fig.\ref{fig:hb}). This arrangement ensured smooth side-wall interaction with minimal energy dissipation. There was no obvious interference between the longitudinal hexbug motion inside the channel and the transversal oscillations. The hexbug had an overall mass of $m=8.7$ g, including the frame and the two springs. At 760 mV, the internal vibrator period was $T_0$=1/98 s. 

The hexbug was observed from above using a camera operated at 1000 fps. The hexbug motion was followed by video-tracking. First, the hexbug forward speed was measured for various power supply voltages in an open channel (without an end-wall).  The results are plotted in Fig.\ref{fig:hb}b. Two distinct regimes were observed. The hexbug stayed in contact with the baseplate and its speed was relatively low below 470 mV. Above this point, however, the vibrations induced free-flight (ballistic) jumps of the hexbug and its speed increased abruptly. All the remaining measurements were carried out in this second regime 
using~760~mV.

In the next stage of the experiment, the channel was closed by a hard wall. The hexbug hitting the end-wall collided through the rigid tip of the plastic frame, causing an instantaneous back-reflection. The distance $x$ measured between the hexbug tip and the end-wall (see Fig.\ref{fig:hb}) was evaluated for each video frame. Instead of moving the end-wall in the laboratory frame, the hexbug was placed onto a moving base-plate that was pulled along the channel axis at constant velocity. In this sense, the end-wall and side walls moved together in the frame of the base-plate. The end-wall and hexbug motion and velocities presented hereafter are expressed in the base-plate frame. The hexbug bouncing motion was recorded for different end-wall velocities (measuring five times 6 seconds at each condition).

\begin{figure}[h]
\includegraphics[width=12cm]{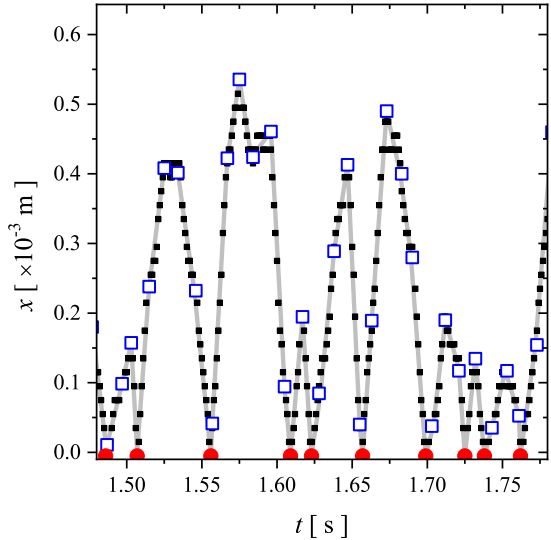}
\caption{\label{fig:timeseries} Typical experimental time-series of the hexbug bouncing motion. The hexbug distance to the end-wall $x$ is plotted as a function of time. The time is measured from the start of the experimental video recording. The end-wall velocity was set to 0.026 ms$^{-1}$. The experimental points (black solid squares)
are connected with straight gray lines to emphasize the ballistic periods. The blue squares indicate
the periodically occurring velocity changing events. Red circles are placed at positions where the
hexbug hits the end-wall.}
\end{figure}

\subsubsection{Evidence for stochastic pulsed propulsion}

Typical time-series of the hexbug distance to the end-wall $x$ are plotted in Fig.\ref{fig:timeseries}. Several details should be highlighted here. First, the amplitude of the bouncing motion fluctuates significantly, indicating that the hexbug dynamics is highly stochastic.  Furthermore, there are distinct time periods when the hexbug moves at a nearly constant velocity, not accelerating or being affected by any noise. A detailed analysis of the trajectories reveals that these longer passive periods are interrupted by very short active events (marked by blue open squares in Fig.\ref{fig:timeseries}) that involve substantial velocity changes. The observed abrupt velocity changes are explained by the short contact of hexbug legs with the base plate. The hexbug vibrator frequency determines the occurrence period of these events. Between velocity changes, the hexbug detaches from the baseplate and moves in ballistic mode.  When projected to the forward-backward direction, as observed by the camera, this ballistic mode appears to be a constant velocity motion. The other type of velocity change occurs when the hexbug collides with and bounces off the end-wall (marked by red solid circles in Fig.\ref{fig:timeseries}).  
All of the facts mentioned above are considered when developing the theoretical model that describes the hexbug bouncing dynamics.

\subsubsection{The system asymmetry} 
\label{asymmetry}
As mentioned in the introduction, the hexbug system has an inherent directional asymmetry. The hexbug legs are bent backward (see Fig.\ref{fig:hb}). Propulsion would not arise in the vibrated hexbug without this asymmetry. Therefore, the asymmetry is critical for making the hexbug active. As shown in the previous section the hexbug velocity changes abruptly when the legs interact with the base-plate. It is natural to assume that this interaction will differ depending on whether the hexbug moves backwards or forwards. The present experimental setup is ideal for investigating these differences. When the end-wall moves quickly against the hexbug propulsion, the hexbug is forced to move backwards for the majority of the time. By contrast, when the end-wall is moved in the direction of the hexbug propulsion, the hexbug moves mostly forward as well. This allows us to gradually shift the experimental conditions from predominantly backward to predominantly forward hexbug motion. If the hexbug dynamics exhibit directional asymmetry, this asymmetry should be reflected in the experimental results obtained at various end-wall velocities, which is the focus of our investigation.
   
\subsection{\label{modelPulse} Pulsed active particle model}

To describe the dynamics of an inertial self-propelled particle striking and bouncing off a rigid wall, a phenomenological stochastic model is proposed. When designing our model, we keep two fundamental goals in mind. First, we focus on an event-driven time pulsed model adapted for the present hexbug toy-robot moving in a narrow channel. The model complexity is tailored to the aspects of motion asymmetry caused by the hexbug specificity. Based on phenomenological considerations, we create a model with two independent parameter sets that distinguish the hexbug velocity direction and allow us to represent differences in active particle forward and backward motion. The second goal is to build a time pulsed model that allows for efficient numerical computations and provides us with spatial and temporal mean values for comparison with the experiment.

\begin{figure}[t]
\includegraphics[width=7cm]{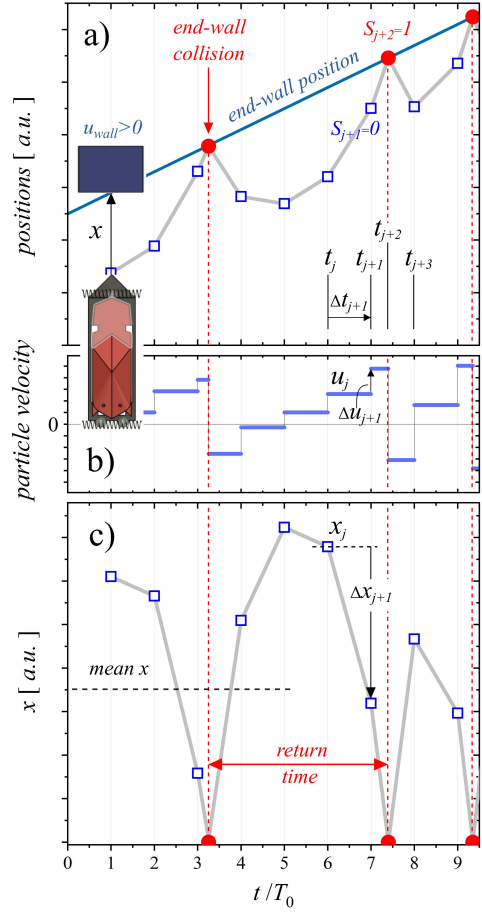}
\caption{\label{fig:model} Schematic illustration of the model assumptions. a) The position of the end-wall moving at a constant velocity (straight blue line) and the position of the particle bouncing against the wall (gray line) are plotted as a function of time. The periodic particle velocity changing events are indicated by blue open squares; the red points show the end-wall collisions. b) The time course of the particle velocity. Constant velocity values belong to ballistic periods. c) The particle distance to the end-wall. The indices of the model time, velocity, and distance to end-wall variables are depicted in the three panels. }
\end{figure}

\subsubsection{The discretized Langevin equation and particle-wall collisions}

The dynamic model of active Brownian particles augmented by particle inertia (see e.g. \cite{OByrne2022, Scholz2018a}) in the limit of one-dimensional motion is the central paradigm in our considerations.
Taking into account the system nature and presumed regimes, we introduce discretized active particle dynamics. The movement under consideration is stochastic and non-linear. Discreteness in the dynamical description is possible when only significant variations in particle velocity exist.
In our model, we use two distinct types of abrupt velocity changes.  The first type of event corresponds to periodic pulses, while the second is caused by particle collisions with the wall.  The particle moves in a ballistic regime between the consecutive velocity changing events. Fig.\ref{fig:model} explains the model assumptions, model nomenclature, and the relationship between the model and the experiments. 
We characterize the particle states at the velocity changing events using quadruples $(t_j, u_j, x_j, S_j)$ which are indexed by the running order $j=0,1,2...$. The first three variables stand for the time $t_j$ when the event occurs, the particle velocity $u_j$ after the event, and the particle distance to end-wall $x_j$ at the time of the event. The quadruple is completed by the discrete binary variable $S_j \in \{0,1\}$. This key variable indicates whether the velocity in step $j$ changes as a result of a periodic pulse ($S_j=0$) or a collision with the wall ($S_j=1$). The value of $S_{j+1}$ is determined by the end-wall velocity $u_{\rm wall}$, and the particle variables at index $j$:

\begin{equation}
S_{j+1}=\mathbf{1}_{
\left[\,0<\frac{x_j}{u_j-u_{\rm wall}}<T_0 \,\,\right]}\,.
\label{eq:IndicsS}
\end{equation} 
\\
The symbol $\mathbf{1}_{[\ldots]}$ in Eq.(\ref{eq:IndicsS}) represents 
the indicator function. Specifically, for the output to be one, both input inequalities must be satisfied; otherwise, the output is zero. The inequalities express whether the particle reaches the end-wall before the next periodic event occurs. 

On a formal level, the behavior of the time, velocity, and distance to end-wall variables ($t_j$, $u_j$, and $x_j$) can be expressed by a set of auxiliary  difference parameters as follows (see also Fig.\ref{fig:model}):
\begin{eqnarray}
\left(
\begin{array}{l} 
t_{j+1} 
\\  
u_{j+1}
\\
x_{j+1}
\end{array}
\right)
&=& 
\left(
\begin{array}{l} 
t_j+\Delta t_{j+1}
\\
u_j+\Delta u_{j+1}
\\ 
x_j+\Delta x_{j+1}
\end{array}
\right)\,.
\label{eq:tujplus1}
\end{eqnarray}
\\
We can reconstruct the behaviour of $x_j$ by repeatedly assigning ${\Delta x}_{j+1} = (u_{\rm wall}-u_{j}) {\Delta t}_{j+1}$. It is important to note, however, that both 
$\Delta t_{j+1}$ and ${\Delta u}_{j+1}$ are affected by the type of event that occurs at $t_{j+1}$. Accordingly, in the upcoming dynamical equations, these parameters are denoted by superscripts p,c  to distinguish whether they correspond to pulses of a periodic nature (superscript p, when $S_{j+1}$=0) or events caused by collisions with the wall (superscript c, when $S_{j+1}$=1).
Using all the facts for 
$S_{j+1}, {\Delta t}_{j+1}, \Delta u_{j+1}$, we get 
the following system of equations:

\begin{equation}
\left(
\begin{array}{l}
\Delta t_{j+1} 
\\
\Delta u_{j+1}
\end{array}
\right)
=
\left(
\begin{array}{l}
\Delta t_{j+1}^{\rm p} 
\\
\Delta u_{j+1}^{\rm p}
\end{array}
\right) (1-S_{j+1})
+
\left(
\begin{array}{l}
\Delta t_{j+1}^{\rm c} 
\\
\Delta u_{j+1}^{\rm c}
\end{array}
\right) S_{j+1}\,.
\label{eq:dtdupc}
\end{equation}
\\
In order to fully understand the peculiarities of $S_j$, 
it is necessary to first consider $\Delta t^{\rm p,c}_{j+1}$ variants.  
The explicit form of time shifts is given by the specifications below:
\begin{equation}
\Delta 
t_{j+1}^{\rm p}= T_0 (1- S_j) + (T_0 -\Delta t^{\rm c}_j)
S_j\,,
\qquad 
\Delta t_{j+1}^{\rm c}=
\frac{x_j}{u_j-u_{\rm wall}}\,.
\label{eq:timeshiftpc}
\end{equation} 
The above use of $S_j$ inside $\Delta 
t_{j+1}^{\rm p}$ implies that the cases $T_0$ and $T_0-\Delta t_j^{\rm c}$  exclude each other.  
For instance, the relation specifies that the preceding end-wall collision decreases the time shift from $T_0$ to $T_0- \Delta t^{\rm c}_j$. 
Naturally, $\Delta t^{\rm c}_{j+1}$ (see Eq.(\ref{eq:timeshiftpc})) is the same term that appears and is compared with $T_0$ in the condition of Eq.(\ref{eq:IndicsS}).
The characterisation of velocity changes completes the dynamics description:
\begin{equation}
\Delta u_{j+1}^{\rm p}=(\,-\gamma^{\rm p} u_j+
f_0^{\rm p}+
\sigma^{\rm p} 
{\mathcal N}_j^{0,1}\,)/m\,,
\qquad 
\Delta u_{j+1}^{\rm c}=
2 (u_{\rm wall}-u_j)\,.
\label{eq:deltau}
\end{equation}
\\
The formula given for $\Delta u_{j+1}^{\rm p}$ is a discretized analogy of the Langevin equation, which defines the active Brownian particle dynamics. It contains damping, propulsion and noise terms with the corresponding parameters denoted by $\gamma^{\rm p}$, $f_0^{\rm p}$ , and $\sigma^{\rm p}$, respectively. The pulsation mode is indicated by the superscript p. All time-dependencies are incorporated within the pulsed parameters in our discrete formulation.  As a result, the physical dimensions of these parameters differ from those of their continuous analogs. For this, the well-known Wiener stochastic coefficient 
$\sim \sqrt{\Delta t_j }$ is already hidden in $\sigma^{\rm p}$. As is standard in stochastic dynamic models, we introduce ${\mathcal{N}}_j^{0,1}$, which is the $j$-th random independent value generated from a Gaussian distribution (with zero mean and unit variance) that is an element of a time series with white noise property. Please keep in mind that in an open channel (without the end-wall) the equilibrated particle velocity will fluctuate around a mean value determined by the propulsion/damping parameter ratio. For simplicity, we will denote this velocity as $u_{\rm max} = f^{\rm p}_0/\gamma^{\rm p}$ and refer to it as the maximal particle velocity.

The changes in particle velocity caused by end-wall collisions are expressed by $\Delta u_{j+1}^{\rm c}$. The formula in Eq.(\ref{eq:deltau}) is consistent with the assumptions that the particle bounces off the solid wall elastically. During the collision, the particle retains its absolute velocity relative to the wall, i.e. $(u_{j+1} - u_{\rm wall}) = -(u_{j} - u_{\rm wall})$. 
It is worth noting that in a collision, the particle direction changes only when the particle travels towards the wall ($u_{j}>0$) and the wall 
velocity is not excessively fast: $u_{\rm wall}<u_{j}/2$. 
In the scenario where $u_{\rm wall} = u_{j}/2$, 
the particle is stopped by the collision ($u_{j+1}=0$).

\subsubsection{Directional asymmetry} 

Two parameter sets are introduced into the model to account for the pronounced directional asymmetry of the experimental hexbug system (see \ref{asymmetry}). The use of the two parameter sets, either $\gamma^{\rm p}_{\rm F}$, $f^{\rm p}_{\rm 0F}$, $\sigma^{\rm p}_{\rm F}$, or $\gamma^{\rm p}_{\rm B}$, $f^{\rm p}_{\rm 0B}$, $\sigma^{\rm p}_{\rm B}$, is conditioned by the particle forward $(F)$ motion ($u_j>0$) or backward $(B)$ motion ($u_j\leq 0$). We express the directional asymmetry also by means of dimensionless asymmetry factors $\alpha_{\gamma}$, $\alpha_{f0}$, $\alpha_{\sigma}$, which represent the ratio of the forward/backward parameter pairs:
\begin{eqnarray}
\alpha_{\gamma}=\frac{\gamma^{\rm p}_{\rm F}}{\gamma^{\rm p}_{\rm B}}\,,
\qquad 
\alpha_{f0} &=&\frac{f^{\rm p}_{\rm 0F}}{f^{\rm p}_{\rm 0B}}\,,
\qquad 
\alpha_{\sigma}=\frac{\sigma^{\rm p}_{\rm F}}{\sigma^{\rm p}_{\rm B}}\,. 
\end{eqnarray}

We need to be able to shift from the asymmetric to the symmetric case for comparison purposes. This requirement is represented by a set of effective (symmetric) parameters $\gamma^{\rm p}_{\rm sym}$, $f^{\rm p}_{\rm 0,sym}$, and $\sigma^{\rm p}_{\rm sym}$, which are the geometric mean of the corresponding forward and backward parameters.: 
\begin{eqnarray}
\begin{array}{ll}
\gamma^{\rm p}_{\rm sym}&=\sqrt{\gamma^{\rm p}_{\rm F} \gamma^{\rm p}_{\rm B}}\,,
\\
f^{\rm p}_{\rm 0,sym} &=\sqrt{f^{\rm p}_{\rm 0F} f^{\rm p}_{\rm 0B}}\,,
\\
\sigma^{\rm p}_{\rm sym}&=\sqrt{\sigma^{\rm p}_{\rm F} \sigma^{\rm p}_{\rm B}}\,.
\end{array}
\end{eqnarray}
\\
If we use the asymmetries 
$\alpha_{\gamma}$, $\alpha_{f0}$, 
and $\alpha_{\sigma}$
then the parametric selection
\begin{eqnarray}
\gamma^{\rm p} &=& \left\{
\begin{array}{llll}
\gamma^{\rm p}_{\rm F} &=& \gamma^{\rm p}_{\rm sym}\sqrt{\alpha_{\gamma}} \,,   &  u_j>0\,\\
\gamma^{\rm p}_{\rm B} &=& \gamma^{\rm p}_{\rm sym}/\sqrt{\alpha_{\gamma}} \,,    &  u_j\leq 0
\end{array} \right.\,\,,
\label{eq:asymFB}
\nonumber
\\
f_0^{\rm p} &=& \left\{
\begin{array}{llll}
 f^{\rm p}_{\rm 0F} &=& f^{\rm p}_{\rm 0,sym}\sqrt{\alpha_{f0}}\,,    &  u_j>0\, \\
 f^{\rm p}_{\rm 0B} &=& f^{\rm p}_{\rm 0,sym}/\sqrt{\alpha_{f0}}\,,    &  u_j\leq 0
\end{array} \right.\,\,,
\\
\sigma^{\rm p} &=& \left\{
\begin{array}{ll}
\sigma^{\rm p}_{\rm F} = \sigma^{\rm p}_{\rm sym}\sqrt{\alpha_{\sigma}} \,,   &  u_j>0\, \\
\sigma^{\rm p}_{\rm B} = \sigma^{\rm p}_{\rm sym}/\sqrt{\alpha_{\sigma}} \,,    &  u_j\leq 0
\end{array} \right.
\nonumber 
\end{eqnarray}
\\
is consistent with the notation used in Eq.(\ref{eq:deltau}).
When the corresponding value of $\alpha$ equals $1$, directional asymmetry disappears for each of these parameter types (damping, propulsion, or noise). 

\subsubsection{Fitting the model parameters to experimental data}\label{section:fit}

The project goal is, among other things, to optimize the model, that is, to identify the parameter region with the best agreement between simulation and experimental results. Since the model is quite complex, and since there is no clear way of solving it analytically, we focus on statistical simulation outputs. For that, a simple numerical scheme is applied to solve the model. The particle distance to the end-wall $x$ is calculated and recorded with a time resolution of $\delta t \,(2.10^{-5}\,{\rm s}) \ll T_0\, (1.10^{-2}\,{\rm s})$ while taking into account all of the event-driven model assumptions defined above. This approach provides us with the necessary high-resolution $x(t)$ data from which the statistical properties are calculated in a straightforward way. In the applied numerical scheme, the exact time of the events (periodic pulses or end-wall collisions) are shifted to the nearest multiple of $\delta t$ with negligible effect on the results.

We process two basic characteristics when analyzing both experimental and simulation data: the equilibrium mean value of the particle distance to the end-wall $\langle x\rangle$ and the mean value of the return time $\langle t_{r}\rangle$, which is the mean of the times required for leaving and returning to the end-wall (see Fig.\ref{fig:model}).
The two means are determined and recorded at selected end-wall velocities 
$u_{{\rm wall},i}\,$; $\,\,$ $i=1,2,$ $\,\ldots,$ $N_{\rm ex}$. In this case, $N_{\rm ex}$ denotes the number of experimental points.    
The experimental data, denoted by (ex), can be arranged into triplets 
$(\,u_{{\rm wall},i}\,$; $\langle x \rangle_{{\rm ex},i}\,; $ $\,\langle t_{r} \rangle_{{\rm ex},i}\,)\,$. 
Similarly, the equivalents generated 
by the simulation can be represented as 
$(\,u_{{\rm wall},i}$; $\,\langle  x  \rangle_{\boldsymbol{\theta}, i}\,$; 
$\langle  t_r \rangle_{\boldsymbol{\theta}, i}\,)$.
In this form, $\boldsymbol{\theta}$ stands for real-valued 6-tuple 
$\boldsymbol{\theta}\equiv$ 
$(\gamma^{\rm p}_{\rm sym}$, $f^{\rm p}_{\rm 0,sym}$, $\sigma^{\rm p}_{\rm sym}$, 
$\alpha_{\gamma}$, $\alpha_{f_0}$, $\alpha_{\sigma})$ 
of independent parameters for which we optimize. 
The details of the fitting procedure are given in Appendix~A.

We used $5 \times 10^7$ simulation time steps $\delta t$
to determine the mean characteristics for each particular combination of parameters
$\boldsymbol{\theta}$ and $u_{{\rm wall},i}$. 
We omitted the first one-sixth of the generated time series to obtain feasible estimates of equilibrium values$\langle \ldots \rangle_{\boldsymbol{\theta},i}$ independent of transients.

\section{Results and Discussion}

\begin{figure}[t]
\includegraphics[width=8cm]{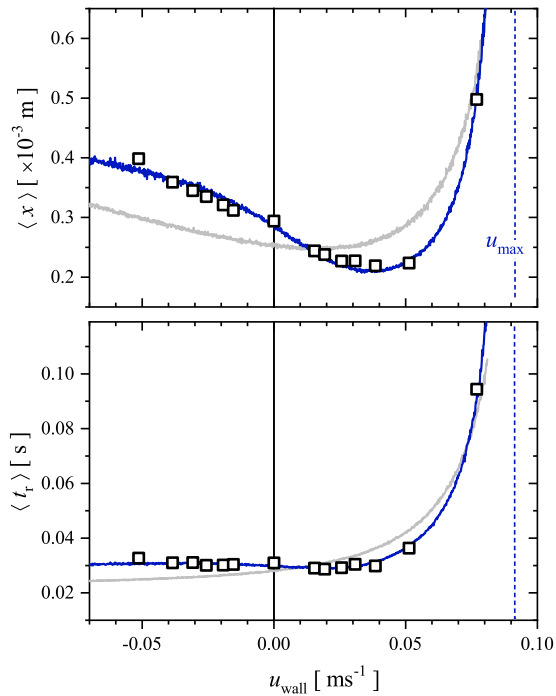}
\caption{\label{fig:xt} a) The mean distance $\langle x\rangle$ between the particle and the end-wall,  and b) the mean return time $\langle t_{r}\rangle$ of the bouncing hexbug/particle plotted against the end-wall velocity. The experimental data are indicated by open square symbols. The optimized simulation results obtained with the symmetric pulsed active particle model (not containing directional asymmetry) are shown by gray solid lines. The blue solid lines represent the results of the optimized asymmetric model. The dashed line indicates the maximal particle velocity calculated from the forward parameters of the asymmetric model: $u_{\rm max} = f^{\rm p}_{\rm 0F}/\gamma^{\rm p}_{\rm F}$.} 
\end{figure}

\subsection{The experimental mean values, $\langle x\rangle_{\rm ex}$ and $\langle t_{r}\rangle_{\rm ex}$  }

The mean values of $\langle x\rangle$ and $\langle t_{r}\rangle$ are plotted against the end-wall velocity in Figure \ref{fig:xt}a and Figure \ref{fig:xt}b, respectively. The details observed in the experimental data that are indicated by open squares in Figure \ref{fig:xt} are addressed in this section.

It can be seen that the minimum of $\langle x \rangle_{\rm ex}(u_{\rm wall})$ does not correspond to the static end-wall case ($u_{\rm wall}=0$). Instead, in terms of the mean $x$ value, the hexbug approaches the end-wall most closely when the wall is moving at a constant velocity $u_{\rm wall}=0.04$~m s~$^{-1}$  in the forward direction of motion, i.e., in the direction of hexbug propulsion. 
This phenomenon arises from the dynamics of end-wall collisions expressed by $\Delta u_{j+1}^{\rm c}$ in Eq.(\ref{eq:deltau}). When the hexbug and the wall are moving in the same direction, the hexbug cannot effectively bounce off the wall. As the end-wall velocity increases further, $u_{\rm wall}$ approaches the maximum velocity $u_{\rm max}$, and the hexbug gradually loses contact with the wall. 
Stochastic forces are important in this process. Without the stochastic forces, the particle could theoretically stay very close to a wall that moves only slightly slower than $u_{\rm max}$. However, because of stochastic fluctuations, the hexbug occasionally bounces off the wall and gets to a larger distance $x$. In this case, when $u_{\rm wall}\lessapprox u_{\rm max}$, it takes a long time for the hexbug to catch up with the wall again. This results in a significant increase of both the mean 
distance $\langle x\rangle_{\rm ex}$ and the mean return time $\langle t_{r}\rangle_{\rm ex}$ 
(see Fig.\ref{fig:xt}) with divergences of 
$\langle x \rangle_{\rm ex}$, $\langle t_r \rangle_{\rm ex}$ at $u_{\rm wall}=u_{\rm max}$.
The situation $u_{\rm wall}<0$ corresponds to the case where the wall moves in the opposite direction of the hexbug propulsion.  
We found amplified reflections from the wall here. As a result, 
there is an increase in 
$\langle x\rangle_{\rm ex}$ on $(-u_{\rm wall})$.

\subsection{Fitting the model to experimental data}

The connections between the experiments and the proposed theory are discussed in detail in this subsection.
Regression results are explained using various levels of model asymmetry to justify adding or eliminating parameters.
The applied parameter fitting method takes both the spatial and temporal mean characteristics $\langle x \rangle, \langle t_r \rangle$ into account. The regression details are described in Appendix A.

\subsubsection{The symmetric model variant}

As a first step, we regressed and tested the basic 
symmetric model represented by a reduced set of parameters 
$\boldsymbol{\theta}_{\rm sym} \equiv $ 
$(\gamma^{\rm p}_{\rm sym}, f_{\rm 0,sym}^{\rm p}$,  
$\sigma_{\rm sym}^{\rm p})$. 
The results are plotted in Fig.\ref{fig:xt} by gray solid lines, the obtained optimized parameters are presented in Table \ref{tab:data1}.

\begin{center}
\begin{table}[h]
\caption{The optimized parameters of the pulsed symmetric model. The listed values are rounded to two significant digits.}
\setlength{\tabcolsep}{5pt} 
\begin{tabular}{ccc}
\hline
\hline
${\gamma^{\rm p}_{\rm sym}}$  & ${f^{\rm p}_{\rm 0,sym}}$     & 
${\sigma^{\rm p}_{\rm sym}}$  
\\ 
\,\,($\times  10^{-3} \mathrm{kg}$) & 
\,\,($\times  10^{-4} \mathrm{kg\, m \,s}^{-1}$)  
& 
\,\,($\times  10^{-5} \mathrm{kg\, m \, s}^{-1}$) 
\\ 
\hline
\hline
1.8 & 1.8  &  8.5  \\
\hline
\end{tabular}
\label{tab:data1}
\end{table}
\end{center}

The symmetric model can only to a limited extent reproduce the main characteristics of the experimental data.  The diverging behavior of  $\langle x\rangle_{\rm ex}$ and 
$\langle t_{r}\rangle_{\rm ex}$ at 
$u_{\rm wall} \rightarrow u_{\rm max}$ is well reproduced, but the overall agreement between the model and the experiment is relatively poor. In general, the model cannot fit both the $\langle x\rangle_{\rm ex}$ and 
$\langle t_{r}\rangle_{\rm ex}$ data well. All of this points to the symmetric model's significant limitations and the need to introduce directional asymmetry into the particle/hexbug description.

\subsubsection{The asymmetric model}

\begin{figure}[h]
\includegraphics[width=16cm]{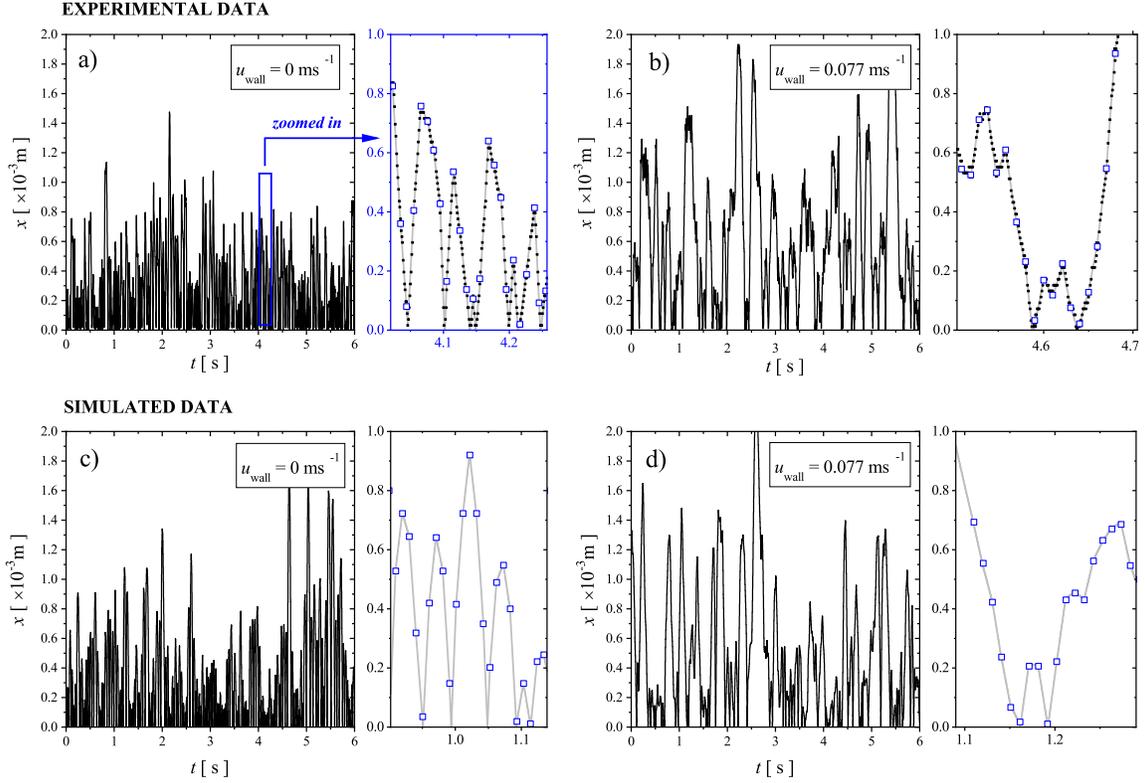}
\caption{\label{fig:rawdata} Typical experimental (a,b) and simulated (c,d) time-series of the hexbug bouncing motion in case of a static (a,c: $u_{\rm wall}=0$ ms$^{-1}$) and moving (b,d: $u_{\rm wall}=0.077$ ms$^{-1}$) end-wall. Zoomed in data are shown on the right side of each panel. The experimental zoomed in points (black solid squares) are connected with straight gray lines to emphasize the ballistic periods. The open blue squares indicate the periodically occurring velocity changing events.}
\end{figure}

The agreement between simulated and experimental results improves significantly in many qualitative aspects when a full pulsed active particle model with directional asymmetry is used. The asymmetric model results are represented by blue solid lines in Fig.\ref{fig:xt}. The optimized $\langle  x  \rangle_{\boldsymbol{\theta}, i}$ values are very close to the experimental $\langle x \rangle_{{\rm ex},i}$ points 
(black open squares in Fig.\ref{fig:xt}a). Moreover, as shown in Fig.\ref{fig:xt}b, the asymmetric model reproduces the return times $\langle t_{r} \rangle_{{\rm ex},i}$ very well. 

Typical sets of raw experimental and simulated data are shown in Fig.\ref{fig:rawdata} to depict the bouncing dynamics stochasticity and the similarity  of the experimental and simulated time-series at two different end-wall velocities.     

\begin{center}
\begin{table}[t]
\caption{
The optimized parameters of the pulsed asymmetric model. 
The listed values are rounded to two significant digits.}
\setlength{\tabcolsep}{5pt} 
\begin{tabular}{cccccc}
\hline
\hline
${\gamma^{\rm p}_{\rm sym}}$ & ${\alpha_{\gamma}}$ & ${f^{\rm p}_{\rm 0,sym}}$ & ${\alpha_{f_0}}$ & ${\sigma^{\rm p}_{\rm sym}}$  & ${\alpha_{\sigma}}$ \\ 
($\times  10^{-3} \mathrm{kg}$) &  & ($\times  10^{-4} \mathrm{kg\, m \,s}^{-1}$) & & ($\times  10^{-5} \mathrm{kg\, m \, s}^{-1}$)\\ \hline
1.4 & 2.3  & 1.9 & 0.99 & 6.7 & 1.4 \\
\hline \hline
\end{tabular}
\label{tab:data}
\end{table}
\end{center}

The model parameters optimized for the experimental hexbug data are given in Tab.\ref{tab:data}. The damping term $\gamma^{\rm p}$ has a significant asymmetry, with enhancement in the forward direction as indicated by $\alpha_{\gamma}>1$. 
The propulsion $f^{\rm p}_0$ is nearly symmetric ($\alpha_{f_0}\approx 1$), whereas the stochastic term has a predominance of forward stochasticity ($\alpha_{\sigma}>1$). 

As discussed in Section \ref{asymmetry}, the hexbug directional asymmetry is caused by the hexbug legs bending backwards. Given the viscoelastic properties of all rubber-like polymer materials (such as those used for hexbug legs), it is reasonable to assume that system damping is largely determined by energy dissipation when the legs are deformed. Intuitively, the extent of these deformations is greater for forward motion, which corresponds well with the pronounced asymmetry identified for the damping term $\gamma^{\rm p}$. The simulations also show that the forward motion is more stochastic. There is no simple explanation for the asymmetry of the stochastic parameter.

\subsection{Spatial, temporal and frequency domain distributions}

  \subsubsection{Distance to end-wall and return time probability distributions}

The $\langle x\rangle_{\rm ex}$ and $\langle t_r\rangle_{\rm ex}$ values given in the previous section represent the integral characteristics of the hexbug dynamics used to optimize the model parameters. The quality of the optimized model in describing hexbug dynamics can be validated by comparing the experimental and simulated probability density functions for the hexbug distance from the end-wall $x$, and the hexbug return time $t_r$. 

\begin{figure}[t]
\includegraphics[width=8cm]{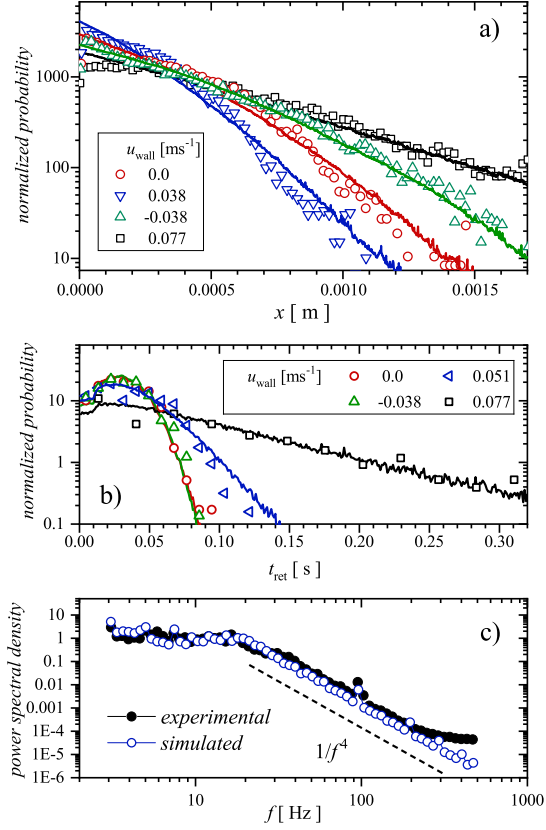}
\caption{\label{fig:distr} a) Spatial probability distributions measured experimentally (open points) and calculated with the optimized asymmetric model (solid lines) at different end-wall velocities. b) Experimental and simulated return time probability distributions. c)~The power spectral density of the hexbug/particle oscillations in a narrow channel closed by a static end-wall. Black solid symbols: experimental data; blue open circles: simulation results. The dashed straight line indicates a $1/f^4$ power function. All the distributions in panels a), b) and c) were normalized such that their integral over the entire range is equal to unity.}
\end{figure}

Fig.\ref{fig:distr}a depicts the experimental and simulated spatial probability distributions for four different end-wall velocities. 
The distribution integrals are normalized to unity. In all cases, the highest probability of finding a hexbug is in the close vicinity of the wall. The experimental distributions (open symbols) exhibit non-exponential decay towards larger distances, which the simulations reproduce well. 

Selected probability distributions of the hexbug return times are plotted in Fig.\ref{fig:distr}b. The experimental data suggest that the most likely return time is around 0.03 s, at least for lower and negative end-wall velocities. Again, the experimental and modeling results are in good agreement.

\subsubsection{Power spectral density function}

The time-related characteristics of the trapped particle dynamics can also be analyzed in the frequency domain. The power spectral density of the oscillating particle position is calculated for the case of a static end-wall ($u_{\rm wall}=0$). The experimental and simulated results are plotted in Fig.\ref{fig:distr}c. The model accurately reproduces the general shape of the experimental power spectral density.  The vibrator frequency appears as a peak in the distribution near 100 Hz, and higher harmonics are also visible in the simulated results. The experimental curve levels off near the Nyquist frequency (500 Hz) due to aliasing \cite{Berg-Sorensen2004}. Both the measured and simulated power spectral densities exhibit an asymptotic $1/f^4$ dependence in the high-frequency domain (above 30 Hz), demonstrating the pronounced importance of the hexbug inertia.  The $1/f^4$ behavior can be explained by comparing the current system to the trapped particles for which the Langevin equation can be solved analytically.  In the simplest case of passive particles submerged in viscous fluids 
and trapped in a harmonic potential 
the high-frequency power spectral density of the Brownian oscillations approaches asymptotically $1/f^2$ and $1/f^4$ for particles with negligible and non-negligible mass, respectively \cite{Berg-Sorensen2004,Lukic2007}. In this consideration, the inertia of the surrounding fluid is omitted, which is a reasonable analogy to the current macroscopic particles moving in air. Despite the fact that we are dealing with a pulsed version of the Langevin equation and that particle trapping is caused by particle propulsion combined with particle bouncing off the end-wall (rather than a harmonic potential), the $1/f^4$ dependence clearly dominates the power spectral distribution at high frequencies.   

\subsection{Characteristics revealed by simulations}

\subsubsection{The relevance of individual force terms}

\begin{figure}[t]
\includegraphics[width=8cm]{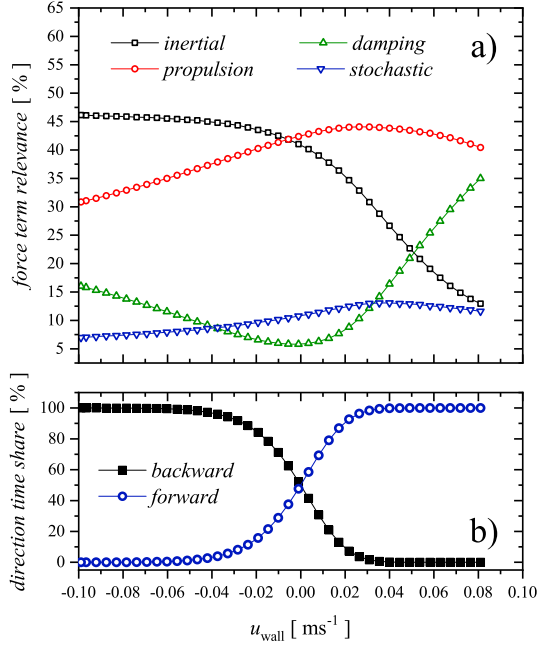}
\caption{\label{fig:forceterms} Simulated data obtained with the optimized asymmetric model. a)~The relative importance of the individual force terms (sub-mechanisms) at different end-wall velocities. The contributions from the forward 
and backward motion are not distinguished. b)~The percentage of time spent 
in forward vs. backward motion. }
\end{figure}

The phenomenological model can provide information that is not available experimentally. Model modifications and relevant simulation techniques can also be used to predict the behavior of other systems and to assess the significance of individual term and parameter values.  In this section, we will examine the relative importance of the various force terms (sub-mechanisms) that contribute to velocity changes.  

To assess the impact of the sub-mechanisms at different $u_{\rm wall}$ we chose to evaluate their mean absolute values in successive velocity changing events.
In doing so, we rely solely on the four terms in the pulsed Langevin equation 
(see the left side of Eq.(\ref{eq:deltau})). We avoid the velocity changes associated with end-wall collisions. 
The relative measure $\delta_k/\sum_{l=1}^4 \delta_l$, 
which is evaluated as a function of $u_{\rm wall}$ in the simulations, 
takes the inertia $\delta_1=\langle\vert m\Delta u^{\rm p}_{j+1} \vert\rangle$, 
damping $\delta_2=\langle\vert \gamma^{\rm p} u_j\vert\rangle$, 
propulsion $\delta_3=\langle\vert f^{\rm p}_0\vert\rangle$ 
and the stochastic $\delta_4=\langle\vert\sigma^{\rm p} \mathcal{N}^{0,1}_j \vert\rangle$ terms into account. 
The obtained results are plotted against the end-wall velocity in Fig.\ref{fig:forceterms}a.

The lower panel in Fig.\ref{fig:forceterms} shows the percentage of total time spent in forward or backward motion (relative to the base-plate). Different dynamic regimes can be identified when the two panels are perceived together. First, when the end-wall velocity $u_{\rm wall}$ approaches the maximal hexbug velocity ($u_{\rm max}\approx0.09$ ms$^{-1}$), forward motion is dominant (see Fig.\ref{fig:forceterms}b). At these conditions, the particle moves at nearly constant speed, the velocity changes are small, and the inertial term is low in Fig.\ref{fig:forceterms}a. The damping term compensates for the propulsion force. Second, forward and backward motion are represented equally in the region of a static or slowly moving end-wall (near $u_{\rm wall}$=0). 
As indicated by a pronounced minimum in the damping term, there are no fast-velocity periods in either direction here. The propulsion is required to stop and re-accelerate particles moving backwards after the end-wall collision. As a result, the propulsion and inertial terms are almost equal. Third, when the end-wall moves in the opposite direction as the propulsion force, the particle is frequently pushed backwards. 
The role of propulsion is reduced in this regime. 

\subsubsection{The effect of asymmetry parameters}

The most interesting findings in this paper are related to the concept of directional asymmetry. The simulations regressed onto the experimental mean values show that asymmetry exists primarily in the damping and stochastic terms, as indicated by the asymmetry parameters $\alpha_{\gamma}$, and $\alpha_{\sigma}$ reported 
in Tab.\ref{tab:data}. The model, however, allows also for the 
asymmetry in the propulsion term $\alpha_{f_0}$, which may be present in other experimental systems.

\begin{figure}[t]
\includegraphics[width=8cm]{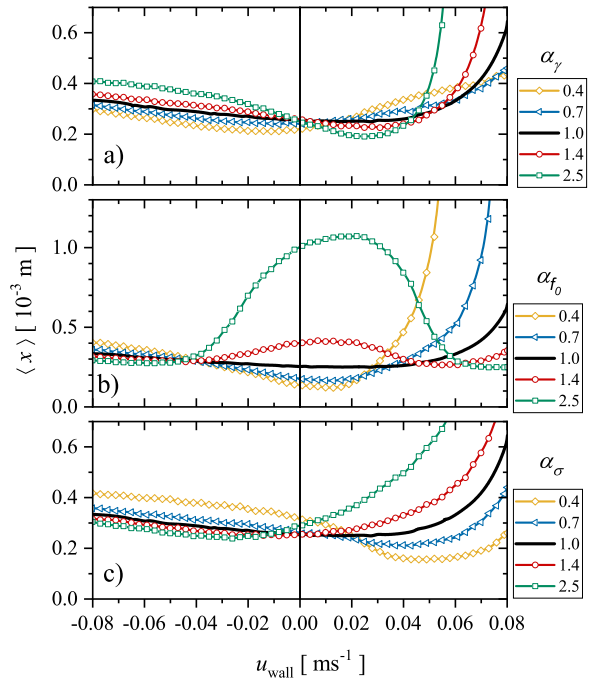}
\caption{\label{fig:modelresults}
The mean distance $\langle x\rangle$ between the particle and the end-wall simulated 
for various asymmetry conditions. The results obtained with the optimized symmetric model (indicated by solid black lines) is used as a reference. Individual asymmetry parameters $\alpha_{\gamma}, \alpha_{f_0}$ and $\alpha_{\sigma}$ are varied in panel a), b) and c), respectively.}  
\end{figure}

The three different types of asymmetry can be investigated separately using particle simulations. We start with the symmetric model variant (see the gray line 
in Fig.\ref{fig:xt}a) and change the asymmetry parameters one-by-one. The results are analyzed in terms of the particle mean distance to the end-wall at different end-wall velocities, as plotted in Fig.\ref{fig:modelresults}. The symmetric situation is 
indicated by black solid lines that are identical across all three panels.  Asymmetry is introduced into the damping, propulsion, and stochastic factors in panels a), b), and c), respectively.

Trivially, the maximal hexbug velocity $u_{\rm max}$ decreases when the forward damping is enhanced ($\alpha_{\gamma}>1$, red circles and green squares in Fig.\ref{fig:modelresults}a) or the forward propulsion is lowered ($\alpha_{f_0}<1$, blue triangles and yellow diamonds in Fig.\ref{fig:modelresults}b). The introduction of asymmetry in the stochastic term has no effect on the value of $u_{\rm max}$ (Fig.\ref{fig:modelresults}c).

In the case of a static end-wall, the asymmetry of the damping and stochastic terms has no effect on the mean distance values (Fig.\ref{fig:modelresults}a,c). By contrast, when the propulsion term is made asymmetric, favoring the forward direction ($\alpha_{f_0}>1$), the particle mean distance to the end-wall increases significantly (Fig.\ref{fig:modelresults}b). In this case, the particle collides with the static wall at high speed. It moves quickly backward after bouncing off the wall. However, because of the low value of $f^{\rm p}_{\rm 0B}$, it takes a long time to stop the particle and change its direction. As a result, the particle drifts away from the wall.  The opposite is true for reverse asymmetry ($\alpha_{f_0}<1$). The particle hits the wall at a low speed, and the backward motion is stopped quickly after the reflection. As a result, the mean distance reduces. 

In general, the three asymmetry parameters have different effects,  creating minima in the $\langle x\rangle$($u_{\rm wall}$) dependencies. Although not covered in this paper, the use of multiple asymmetries at the same time can result in a significant increase in the output complexity.

\section{Conclusions}
Many previous works have investigated the dynamics and collective motion of inertial active particles. In most cases, particles were allowed to move predominantly forward as defined by their propulsion.  The particle interaction with obstacles and/or other particles may result in situations in which the particle is forced to move against its internal propulsion, at least for short periods of time.  This situation was studied in details in this work. 
The one-dimensional dynamics of modified self-propelled vibrated toy-robots (hexbugs) was analyzed in a narrow channel suppressing particle rotations. The channel end was closed by a solid wall, from which the hexbug bounced back. The end-wall was moved along the channel at different velocities, changing the dominant direction of hexbug motion. 

Based on the mechanical design of the hexbug (six pairs of bent legs), significant directional asymmetry between forward and backward motion can be assumed. The results of the experiments and simulations presented in this paper support this hypothesis.  Two experimental integral characteristics, the mean distance to the end-wall and the mean return time, were well reproduced by the proposed pulsed Active Particle model when directional asymmetry was taken into account. Two distinct parameter sets were used in the Langevin equation for forward and backward motions, respectively.  

Understanding directional asymmetry in the context of toy-robots, we believe, can be applied to the dynamics of other groups of active particles.  The qualitative matching of the essential motion components, as well as the encouraging regression results, support this hypothesis. Asymmetry should be always considered if active particles are forced to move in the opposite direction of their propulsion. Our findings might also serve as an impulse for active matter on microscopic scales.

\section*{Acknowledgement}

This work was supported by the Slovak Research and Development Agency through the project APVV-21-0333 and by the grant agency of the Ministry of Education, Science, Research, and Sport of the Slovak Republic (grant no. VEGA 2/0101/22). This publication is the result of the implementation of the project OPENMED (Open Scientific Community for Modern Interdisciplinary Research in Medicine) ITMS2014+: 313011V455 from the Operational Program Integrated Infrastructure funded by the ERDF.

\appendix

\section{Model versus experiment quantification}
\label{app1}

As mentioned in Section \ref{section:fit} 
the optimization regression 
is performed for a
$6-$tuple of independent parameters 
$\boldsymbol{\theta}\equiv (\gamma^{\rm p}_{\rm sym}$, 
$f^{\rm p}_{\rm 0,sym}$, $\sigma^{\rm p}_{\rm sym}$, $\alpha_{\gamma}$, $\alpha_{f_0}$, $\alpha_{\sigma}$).
The proposed regression approach is based on the inclusion of spatial ($\langle x \rangle$) 
and temporal ($\langle t_r \rangle$) 
perspectives, 
which are ultimately reflected in a pair of measures:
\begin{eqnarray}
R_x(\boldsymbol{\theta}) &=& 
\sqrt{
\frac{1}{N_{\rm ex}}\sum_{i=1}^{N_{\rm ex}} 
\left( 
\frac{ \langle x \rangle_{(\boldsymbol{\theta}, u_{{\rm wall},i})} -  
\langle x \rangle_{{\rm ex},i}}{ 
\frac{1}{N_{\rm ex}} 
\sum_{k=1}^{N_{\rm ex}} \langle x \rangle_{{\rm ex},k}} 
\right)^2}  \,\,,
\label{eq:ApOpt1}
\\
R_{t_r}(\boldsymbol{\theta}) &=& 
\sqrt{
\frac{1}{N_{\rm ex}}\sum_{i=1}^{N_{\rm ex}} 
\left( 
\frac{ \langle t_{r} \rangle_{(\boldsymbol{\theta}, u_{{\rm wall},i})} -  
\langle t_{r} \rangle_{{\rm ex},i}}{ 
\frac{1}{N_{\rm ex}} \sum_{k=1}^{N_{\rm ex}} 
\langle t_{r} \rangle_{{\rm ex},k}} 
\right)^2}\,\,.
\label{eq:ApOpt2}
\end{eqnarray}
We express the need to eliminate differences in spatial and temporal units in these measures. 
Thus, normalization factors are used in the denominators.
The relations include $_{{\rm ex},i}$-indexed terms (where $i=1,2,\ldots, N_{\rm ex}$) that correspond to the measure for the $i$-th choice of $u_{\rm wall}$, denoted by $u_{{\rm wall},i}$.
The equally weighted sum of the previous measures is used to calculate the final scalarized objective function (see e.g. \cite{Miettinen2002}), which takes into account both the matching of simulated and experimentally confirmed wall distance values and the matching of simulated and experimental return times

\begin{equation}
R_{x,t_r}(\boldsymbol{\theta})= 
R_x (\boldsymbol{\theta}) + R_{t_r}(\boldsymbol{\theta}) .
\label{eq:ApOpt3}
\end{equation} 
Then, formally the optimal 
\begin{equation}
\boldsymbol{\theta}^{\ast}= 
\mbox{arg} \,
\min_{\boldsymbol{\theta}\in \mathbb{R}^6} 
R_{x,t_r}(\boldsymbol{\theta})
\label{eq:ApOpt4}
\end{equation}
may be proposed as a means of comparing the simulation and experiment.
We ran several hundred iterations of short simulated annealing cycles for the numerical optimization scenario. Simulated annealing is a particularly powerful strategy among various heuristic approaches because it conveniently incorporates randomness into the search 
\cite{LEE201937}. 
The purpose of frequent repetition of the annealing cycle 
was not only to avoid getting stuck in the local minimum, but also to achieve diversity in the selection of 
metaparameters. To gain insight into the robustness of the minimum in the parameterization of the stochastic problem, annealing metaparameters such as the effective temperature (controlling for the acceptance of random trials), as well as the rate and progression of its decay, were varied along with the initial conditions of the simulation.

\bibliography{ezaz}

\end{document}